\documentclass{PoS}

\newcommand\aap{\emph{A\&A}}

\newcommand\mnras{\emph{MNRAS}}
\newcommand\apj{\emph{ApJ}}
\newcommand\apjl{\emph{ApJ}}

\title{The cosmic origin of carbon and manganese}

\ShortTitle{The cosmic origin of carbon and manganese}

\author{\speaker{Thomas Bensby}\\%\thanks{A footnote may follow.}\\
        European Southern Observatory, Alonso de Cordova 3107, Vitacura, Santiago, Chile\\
        E-mail: \email{tbensby@eso.org}}

\author{Sofia Feltzing\\
        Lund Observatory, Box 43, SE-22100 Lund, Sweden\\
        E-mail: \email{sofia@astro.lu.se}}

\abstract{We have determined carbon abundances for 51 dwarf stars
and manganese abundances for 95 dwarf 
stars in two distinct and well defined stellar populations - the 
Galactic thin and thick disks. As these two populations have different chemical 
histories we have been able to, through a differential abundance analysis 
using high-resolution spectra, constrain the formation sites for carbon and 
manganese in the Galactic disk(s).

The analysis of carbon is based on the forbidden [C\,{\sc i}] line at
872.7\,nm which is an abundance indicator that is insensitive to errors
in the stellar atmosphere parameters. Combining these data with our previously
published oxygen abundances, based on the forbidden [O\,{\sc i}] line at
630.0\,nm, we can form very robust [C/O] ratios
that we then used to investigate the origin of carbon and the chemical
evolution of the Galactic thin and thick disks.
We find that the [C/Fe] versus [Fe/H] abundance trends for the thin and 
thick disks are totally merged and being flat for sub-solar metallicities. 
The thin disk that extends to higher metallicities then shows a shallow
decline in [C/Fe] from $\rm [Fe/H]\approx0$ and up to
$\rm[Fe/H]\approx+0.4$. The [C/O] versus [O/H] trends are well separated
between the two disks (due to differences in the oxygen abundances)
and bear great resemblance with the [Fe/O] versus [O/H] trends.
Our interpretation of our abundance trends is that the sources that are
responsible for the carbon enrichment in the Galactic thin and thick
disks have operated on a time-scale very similar to those that are
responsible for the Fe and Y enrichment (i.e., SN\,Ia and AGB stars,
respectively).

For manganese, when comparing our Mn abundances with O abundances for
the same stars we find that the abundance
trends in the stars with kinematics typical of the thick disk can be 
explained by metallicity dependent yields from SN II.
Furthermore, the [Mn/O] versus [O/H] trend in the halo is flat. 
We conclude that the simplest interpretation of our data is that 
manganese most likely is produced in SN II and that the Mn yields for 
such SNae must be metallicity dependent.
}

\FullConference{10th Symposium on Nuclei in the Cosmos\\
		 July 27 - August 1 2008\\
		 Mackinac Island, Michigan, USA}

\begin{document}

%==============================================================================
\section{Stellar sample, observations, and stellar parameters}

%------------------------------------------------------------------------------
\begin{figure}[b] 
\centering
\resizebox{0.55\hsize}{!}{%\includegraphics{FIG_4.eps}
\includegraphics[width=5cm,height=3cm]{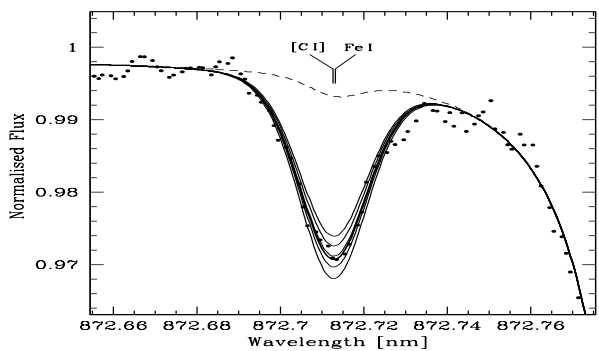}}
\caption{
        The [C\,{\sc i}] line in the solar spectrum. The dots are
        the observed spectrum and the {\bf thick} 
        solid line the best
        fit representing a solar carbon abundance of  $\rm \log
        \epsilon_{\odot}(C) = 8.41$. The thin solid lines represents 5
        different carbon abundances from 8.34 to 8.46\,dex in steps of
        0.03\,dex. The dashed line shows the contribution from the
        blending Fe\,{\sc i} line. The wavelength positions of the
        [C\,{\sc i}] and the Fe\,{\sc i} lines are also indicated.  }
\label{fig:sun_c8727}
\end{figure}
%------------------------------------------------------------------------------

The stellar samples are subsets of the in total 102 F and G dwarf 
stars we presented in \cite{bensby2003,bensby2005}.
On the basis of their kinematic properties the stars have been classified 
to be likely members of either the thin or the thick Galactic disk.
For a detailed description of the kinematical selection criteria
and the kinematical properties of the stars we refer the reader
to \cite{bensby2003,bensby2005,bensby2006}.

High-resolution spectroscopic observations were carried out with three different
spectrographs. First, 69 stars were observed with FEROS in 2000 and 2001
when it was mounted on the ESO 1.5-m telescope on La Silla. Second, another 33 stars
were observed with SOFIN on the Nordic Optical Telescope on La Palma in 2002.
These FEROS and SOFIN spectra with their wide spectral coverage were used
to determine the stellar parameters, as well as for the
determination of Mn abundances. Third, a subset of 51 stars already observed
with FEROS and SOFIN were observed in 2004 with the CES spectrograph on 
the ESO 3.6-m telescope on La Silla. These spectra with their very limited spectral
coverage (only $\sim40$\,{\AA}) but extremely high resolution
($R\approx230\,000$) and high signal-to-noise ($S/N>350$) were used to
analyse the forbidden carbon line, [C\,{\sc i}], at 8727\,{\AA}.

We use the one-dimensional, plane-parallel, LTE,  
Uppsala {\sc marcs} stellar model atmospheres
\cite{gustafsson1975, edvardsson1993, asplund1997}.   Surface
gravities ($\log g$) were determined from Hipparcos parallaxes,
effective temperatures ($T_{\rm eff}$) were determined by requiring
that the abundances derived from  Fe\,{\sc i} lines with different excitation
energies should all yield the same [Fe/H],
and the microturbulence parameter ($\xi_{\rm t}$) by requiring
all Fe\,{\sc i} lines should yield the same
abundances independent of line strength ($\log W_{\lambda}/\lambda$).
All parameters were taken from our previous studies 
\cite{bensby2003, bensby2005} wherein the iterative process to tune 
the stellar parameters also is fully described.

%==============================================================================
\section{Carbon}

The forbidden carbon line is located in the left wing of a Si\,{\sc i}
line and is blended by an Fe\,{\sc i} line.  
The contribution from this blending line to the
joint [C\,{\sc i}]-Fe\,{\sc i} line profile is generally negligible at
sub-solar  metallicities while it in the Sun is estimated to be
between 0.1\,pm and 0.5\,pm \cite{allendeprieto2002}, and will grow in
size for stars with super-solar [Fe/H]. 
Figure~\ref{fig:sun_c8727} shows the observed and the
synthetic spectra for the Sun. From our analysis of the solar spectrum
we get a solar carbon
abundance  of $\rm \log \epsilon_{\odot}(C)=8.41$  when the blending
Fe\,{\sc i} line is taken into account and $\rm \log
\epsilon_{\odot}(C) = 8.44$ when it is neglected. Both are in
good agreement with the recent analysis by \cite{allendeprieto2002,asplund2005} 
who found  a best value of $\rm \log
\epsilon_{\odot}(C) = 8.39$ using 3-D stellar atmosphere models.

%------------------------------------------------------------------------------
\begin{figure}
\centering
\resizebox{0.9\hsize}{!}{ 
 \includegraphics{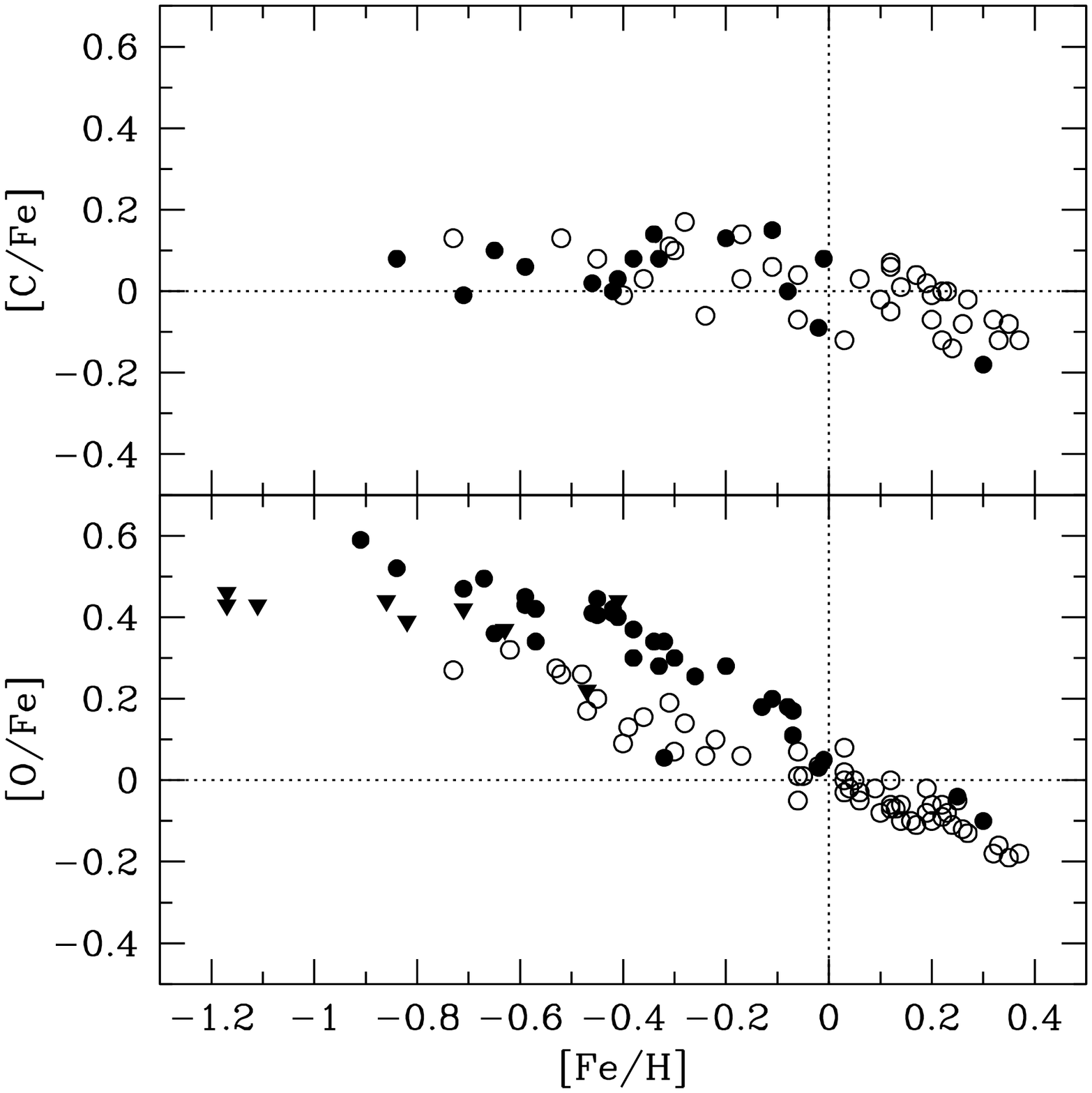}
 \includegraphics{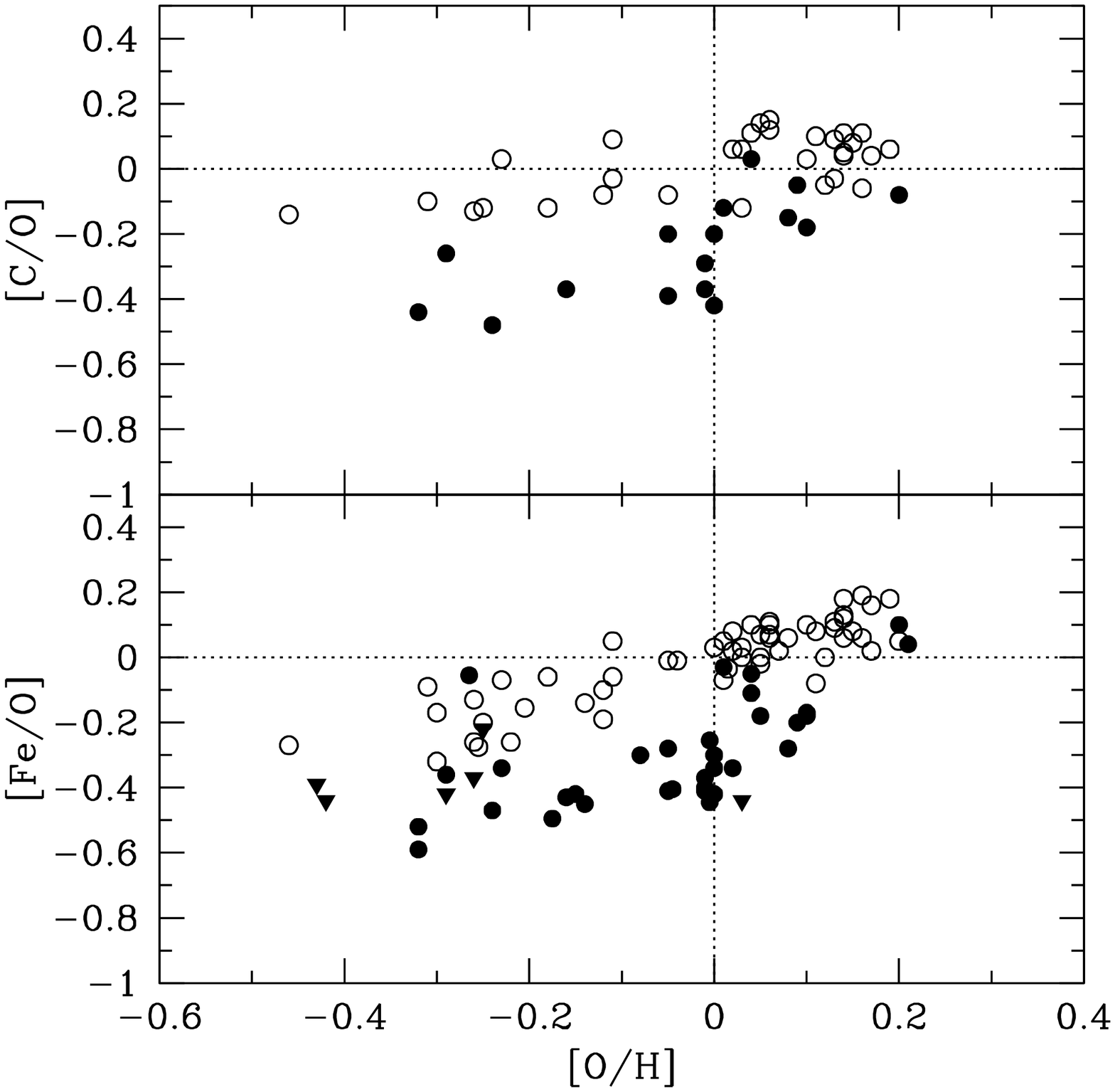}
 }
  \caption{
Our carbon trends relative to iron ({\sl left}) and relative to 
oxygen ({\sl middle}). For 
comparison we also show the [O/Fe] vs [Fe/H] and [Fe/O] vs [O/H]
trends. Thin and thick disk stars are
marked by open and filled symbols, respectively. 
Filled triangles in the oxygen plots denote thick disks stars from
\cite{nissen2002}. 
    }
\label{fig:ctrend}
\end{figure}
%------------------------------------------------------------------------------

%

The resulting abundance trends are shown in Fig.~\ref{fig:ctrend}.
[C/Fe] versus [Fe/H] for the thin and thick disks 
are fully merged and at sub-solar metallicities the [C/Fe] values 
agree within $\rm 0
\lesssim [C/Fe] \lesssim 0.2$ with no particular slope. For 
$\rm [Fe/H]>0$  the [C/Fe] values decrease with increasing [Fe/H].
The [C/O] versus [O/H] trends for the thin and thick disks are 
clearly separated.  The thin disk shows a shallow increase
in [C/O] with [O/H]  whilst the thick disk first has a flat [C/O]
trend that increases sharply at  $\rm [O/H]=0$. The great resemblance
with the observed [Fe/O] versus [O/H] trends also shown in
Fig.~\ref{fig:ctrend} indicates that C and Fe originate
from objects that evolve on similar time scales. 

Our thick disk [O/Fe] trend shows a constant over-abundance of oxygen 
at low [Fe/H] and clear break and subsequent downturn 
at $\rm [Fe/H] \approx -0.4$, a signature that normally is interpreted 
as the onset of SN\,Ia \cite{tinsley1979}. Our [C/Fe] show no such 
trend, instead [C/Fe] have roughly solar values for $\rm [Fe/H] < 0$. 
Hence, we may infer that carbon enrichment happens on the same time scale
as the enrichment from SN\,Ia. Hence the increase in Fe production is
matched by enrichment of C. This is also supported by \cite{chiappini2003b} 
who predict, if low and intermediate mass stars are important, that there 
first would be a flat [C/O] plateau for the thick disk that then 
(at roughly solar [O/H]) should sharply increase, to be followed by a 
shallow thin disk [C/O] trend at higher [C/O] ratios. 

%------------------------------------------------------------------------------
\begin{figure}
\centering
\resizebox{0.95\hsize}{!}{ 
 \includegraphics[bb=18 420 592 700,clip]{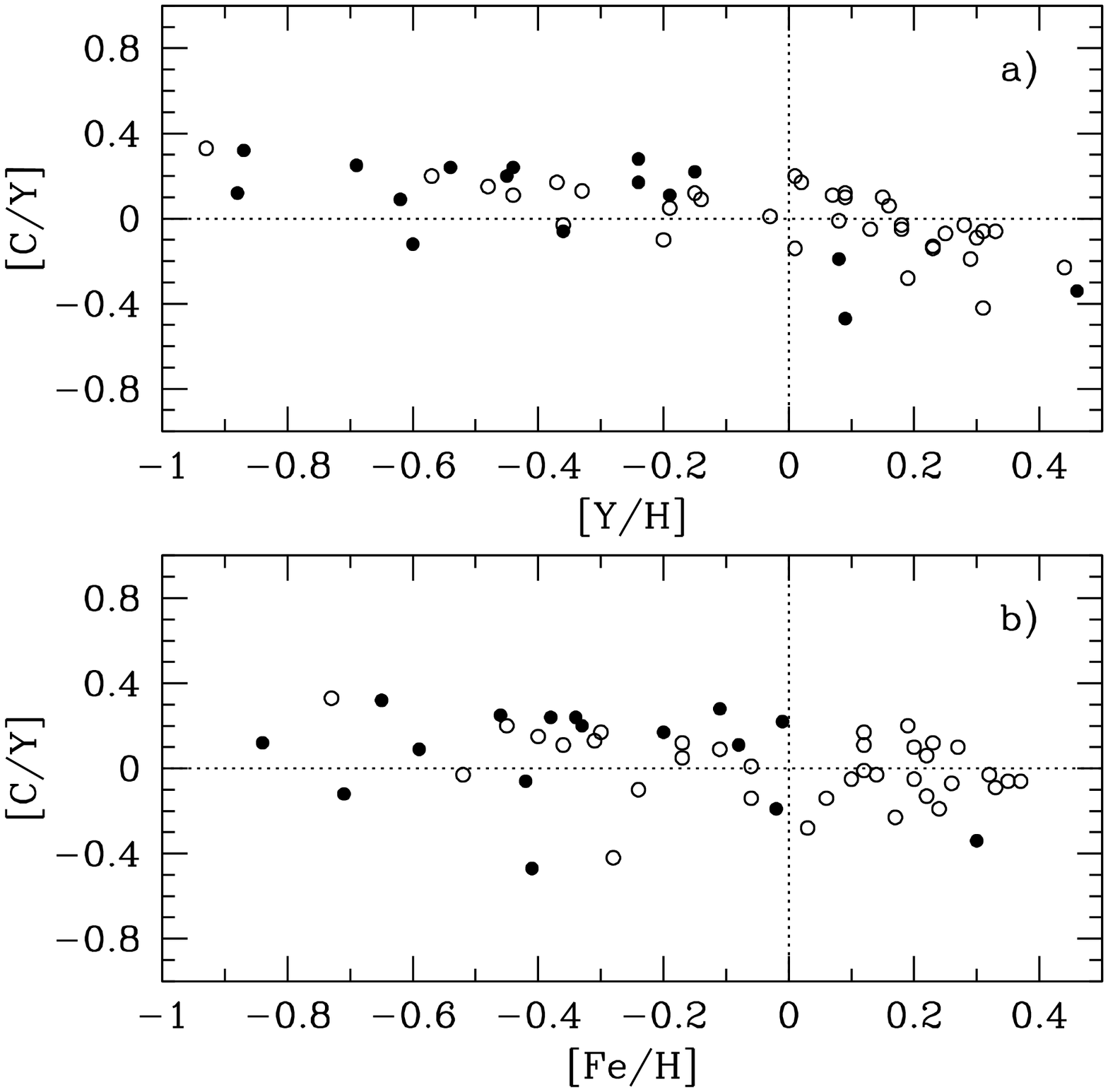}
 \includegraphics[bb=18 150 592 420,clip]{FIG_13.eps}
 }
  \caption{Carbon and yttrium abundance trends. Symbols as in Fig.~2.
    }
\label{ytrend}
\end{figure}
%------------------------------------------------------------------------------
Furthermore, the  trend of 
[C/Y] versus [Y/H] (Y abundances taken from \cite{bensby2005})
shows a flat trend for $\rm [Y/H]\lesssim 0$ which 
turns to a steadily declining trend for $\rm [Y/H]\gtrsim 0$
which is very similar to what we see for [C/Fe] versus [Fe/H] (see
Fig.~\ref{fig:ctrend}).  [C/Y] vs [Fe/H] is flat for both the 
thin and the thick disk samples at all metallicities.  
As Y is mainly produced in the s-process in AGB stars
\cite{travaglio2004}) the most straightforward interpretation of the
lack of trends is that C and Y are made in
objects that enrich the interstellar medium on the same time
scale and that their major components  are
indeed made in the same objects, namely low and intermediate mass
stars in the AGB phase.  

In light of our own as well
as other studies in the literature
\cite{gustafsson1999, chiappini2003, chiappini2003b, 
akerman2004, carigi2005} we feel that the source(s) of carbon is not
yet settled but that there is growing evidence that a complicated,
and finely tuned, set of objects contribute to the enrichment of carbon
in galaxies. As we discussed in \cite{bensby2006}, based on our own 
results only we would conclude that the main source for carbon in the 
Galaxy is low and intermediate mass stars.
However, it appears that massive stars played a
significant r\^ole for the carbon enrichment at low metallicities
(i.e. halo and metal-poor thick disk) whereas low and intermediate 
mass stars dominate more and more at higher metallicities, i.e. that 
they have been the major contributors to the
carbon enrichment in the thin disk and the metal-rich thick disk.

%==============================================================================
\section{Manganese}

\begin{figure}
\centering
\resizebox{0.7\hsize}{!}{\includegraphics[bb=18 480 592 695,clip]{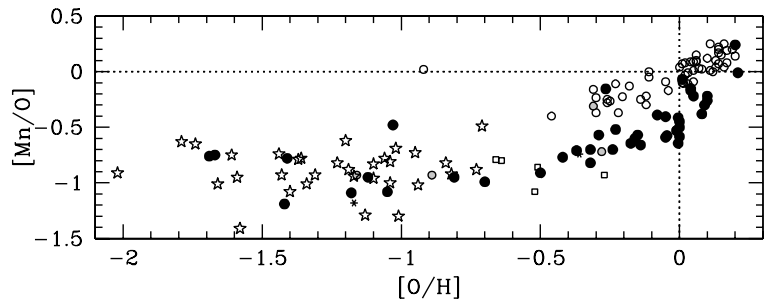}}
\caption{[Mn/O] vs [O/H]. 
$\circ$, $\bullet$, and open stars, indicate thin, thick disk, and halo stars, respectively.}
\label{fig:mn}
\end{figure}

We derive Mn abundances through spectral synthesis of four
Mn\,{\sc i} lines at 539.4, 549.2, 601.3, and 601.6\,nm,
taking the hyperfine structure splitting of the lines into account
(see \cite{feltzing2007} for details).
In order to study the origin of Mn we have combined our new Mn
abundances with oxygen abundances. Fe is made both in SN\,II and in
SN\,Ia. By using oxygen, which is only made in SN\,II, as the reference
element we simplify the interpretation of the abundance data. For our
stars we took the oxygen abundances from our analysis of the forbidden
oxygen line at 630.0\,nm \cite{bensby2004,bensby2005}.
and added data from a number of other studies of (mainly) giant stars 
in the disks and halo of the Milky Way (see \cite{feltzing2007} for 
references to the different sources).
Figure~\ref{fig:mn} shows the trend of [Mn/O] versus [O/H].
For the stars with kinematics typical of
the thin disk we see a steady increase in [Mn/O] as [O/H]
increases. Stars with kinematics typical of the thick disk show a 
similar trend, albeit with an offset of $\sim 0.3$\,dex, for 
$\rm [O/H]<0$. For higher [O/H] there 
is a hint of a faster increase in [Mn/O].

%------------------------------------------------------------------------------
%------------------------------------------------------------------------------

For the halo and metal-poor thick disk, $\rm [O/H]\leq -0.5$, the
[Mn/O] trend is flat.  This indicates that the production of Mn and O
are well balanced. Moreover, we know from the study of 
\cite{bensby2004} 
that the archetypal signature of SN\,Ia in the thick disk
do not occur until [O/H] = 0. Hence the up-going trend we see after
[O/H] $\simeq -0.5$ must be interpreted as being due to metallicity
dependent Mn yields in SN\,II. The rising trend seen for the thin disk
sample could also be interpreted in this fashion. Although here we do
know that SN\,Ia contribute to the chemical enrichment and hence the
increase might also be due to these objects.
Our interpretation is that these data, to first order, can be
explained by metallicity dependent yields in SN\,II. This is,
essentially, in agreement with the conclusions in 
\cite{mcwilliam2003}.

%==============================================================================
\section{Summary}

We have used a two well-defined samples of thin and thick disk stars, both with their
own chemical histories, to put further constraints on the cosmic production sites
of carbon and manganese. Our findings indicate that carbon in the Milky Way disk
mainly comes from low and intermediate mass stars (SN\,Ia and AGB stars) while
manganese is likely to depend on metallicity dependent yields from massive stars.
For further details regarding the analysis methods and a full discussion of 
the results we refer the reader to Bensby \& Feltzing~(2006) and Feltzing et al.~(2007).

%==============================================================================

\end{document}